\documentstyle[11pt,epsf]{article}
\input psfig
\def\beq{\begin{eqnarray}}
\def\eeq{\end{eqnarray}}
\def\bea{\begin{eqnarray*}}
\def\eea{\end{eqnarray*}}

\def\centeron#1#2{{\setbox0=\hbox{#1}\setbox1=\hbox{#2}\ifdim
\wd1>\wd0\kern.5\wd1\kern-.5\wd0\fi
\copy0\kern-.5\wd0\kern-.5\wd1\copy1\ifdim\wd0>\wd1
\kern.5\wd0\kern-.5\wd1\fi}}
\def\ltap{\;\centeron{\raise.35ex\hbox{$<$}}{\lower.65ex\hbox{$\sim$}}\;}
\def\gtap{\;\centeron{\raise.35ex\hbox{$>$}}{\lower.65ex\hbox{$\sim$}}\;}

\def\singleandthirdspaced{\baselineskip=\normalbaselineskip\multiply
    \baselineskip by 130\divide\baselineskip by 100}

\newcommand{\newc}{\newcommand}
\newc{\qbar}{{\overline q}}
\newc{\Kahler}{K\"ahler }
\newc{\deltaGS}{\delta_{\rm GS}}
\begin{document}
\begin{titlepage}
\begin{flushright}
{\large hep-th/0402101 \\ SCIPP-2004/12\\
}
\end{flushright}

\vskip 1.2cm

\begin{center}

{\LARGE\bf Is There a String Theory Landscape:  Some Cautionary Remarks}

\vskip 1.4cm

{\large  Michael Dine}
\\
\vskip 0.4cm
{\it Santa Cruz Institute for Particle Physics,
     Santa Cruz CA 95064  } \\

\vskip 4pt

\vskip 1.5cm

\begin{abstract}
There is evidence that string theory possesses a 
large discretuum of stable and/or metastable ground states, with 
zero or four supersymmetries in four dimensions.  I discuss 
critically the nature of this evidence.  Assuming this 
``landscape" exists, anthropic explanations of some quantities 
are almost inevitable.  I explain that 
this landscape is likely to lead to a prediction of low energy 
supersymmetry.  But we argue that many features of low energy 
physics are not anthropic and, as currently understood, the 
landscape picture will get them wrong. This indicates that this 
viewpoint is potentially falsifiable. Moreover, if it is correct, 
many questions must be answered through more conventional 
scientific explanations. This is based on talks presented at
the conference QTS3 at the University of Cincinnati and at the
KITP conference on Superstring Cosmology in 2003.
\end{abstract}

\end{center}

\vskip 1.0 cm

\end{titlepage}
\setcounter{footnote}{0} \setcounter{page}{2}
\setcounter{section}{0} \setcounter{subsection}{0}
\setcounter{subsubsection}{0}

\singleandthirdspaced


\section{Introduction:  The ``Vacuum Selection" Problem}

From the beginning, the seemingly
vast array of possible ground states
has made string theory both attractive 
and problematic.   Ground states with more than four supersymmetries
have the virtue that they are theoretically tractable,
but they are also totally unrealistic.  It has long 
been clear that no potential for the moduli exists, and the duality 
revolution spoiled any remaining hope that some sort of 
non-perturbative inconsistency might permit us to discard these 
states.  It also strongly suggested that this proliferation of 
possible ground states is an inherent feature of any sort of 
quantum general relativity.  Apart from anthropic arguments (to 
be discussed below), we have no inkling why nature doesn't select 
one of these states.  With four or less supersymmetries there is 
a vast proliferation of candidate ground states, revealed in 
various approximations.  Some of these have features which
resemble those of the real world.  Unlike the case of more supersymmetries, 
there are potentials for the moduli, tadpoles (either at the 
perturbative or non-perturbative level), and some possibility of 
non-perturbative anomalies.  Faced with this plethora
of states, I, for a long time, comforted myself 
that not a single example of a (meta)stable ground state of this 
sort had been exhibited in a controlled approximation, and so 
perhaps there might be some unique or at least limited set of 
sensible states.

One of the most exciting -- and troubling -- developments in 
string theory in the last few years has been the suggestion that 
there is a vast array of stable or highly metastable states of 
string theory with four or less supersymmetries. Crucial to the 
emerging picture is the role of compactification with 
fluxes.\cite{bp,sethietal,becker,sethib,gkp,acharya} The most 
persuasive elaboration of this possibility to date is due to 
Kachru, Kallosh, Linde and Trivedi (KKLT),\cite{kklt} who argue for the 
existence of a {\it discretuum} or {\it 
landscape},\cite{landscape} both supersymmetric with $N=1$ 
supersymmetry, as well as non-supersymmetric, with supersymmetry 
softly broken.
The existence of a landscape, if established, raises questions 
about the very nature of scientific explanation.  Most 
importantly, this assertion places the anthropic principle at 
center stage. There has been strong reaction to this fact, 
ranging from near celebration by advocates of the anthropic 
principle to a great deal of handwringing and even denunciation 
from those who find the anthropic principle objectionable.

In this talk, I would like to give an overview of some of the 
issues raised by the possible existence of a landscape.  I will 
explain why, even before we accept the landscape, some element of 
anthropic explanation is probably inevitable in quantum general 
relativity. Understanding the number of supersymmetries and the
dimension of space-time may well require invoking some
{\it extremely} weak anthropic considerations (what we might
call the Minimalist Anthropic Principle, or MAP).

But the landscape requires a much broader application of the anthropic
principle.  I will stress in this talk is that the questions of
the applicability and validity, of the anthropic principle
are {\it scientific} ones.  As to applicability, I will discuss
what I believe has been reliably established by KKLT, and what
has not (in this context I will present a few, admittedly very
tenuous reasons, to hope that some sort of unique or nearly unique
set of predictions emerge from string theory).
I will mention various proposals for anthropic 
explanations of the cosmological constant, and show that, apart 
from the landscape, our experience with string theory renders the 
others (almost) implausible. 

Then I will turn to the question of whether the anthropic principle
is predictive or falsifiable.  
Michael Douglas\cite{douglas} has put forward, most clearly and 
persuasively, the question of anthropic prediction in this 
context, though he prefers not to use this language.  In 
particular, he stresses that the important question is 
determining the probability distributions for various physical 
quantities in the flux discretuum.  One could imagine that, if 
these distributions were well understood, the anthropic principal 
could become predictive -- and falsifiable.  The typical vacuum 
consistent with anthropic constraints might not look at all like 
the Standard Model, or it might make a prediction, say of 
supersymmetry or some large dimension.  Indeed, while these 
distributions are not yet sufficiently well known to make 
definite statements, I will argue that it is almost inevitable 
that this framework will predict low 
energy supersymmetry.  But I will also explain why it is also likely
to make a number of incorrect predictions.   More 
precisely, if the flux discretuum is to describe nature, there 
are many questions whose explanations cannot be anthropic, but 
must emerge as a result of physical principles.  These principles
are not apparent in the landscape as envisioned by KKLT.  This may be a 
sign that the discretuum is the {\it wrong} direction 
to understand fundamental issues in physics.  Alternatively, it 
may mean that the business of particle physicists will be 
extracting these principles.  In either case, we should neither 
{\it despair nor give up!}

\section{How Many Supersymmetries?}

Why don't we live in a universe with more than four 
supersymmetries?   As I indicated above, there is almost 
certainly nothing wrong with these vacua.  One could imagine that 
these are somehow disjoint theories from that which describes 
what we observe.
This viewpoint has been put forward by Banks.\cite{bankscritique}
It has long seemed to me that some very mild anthropic 
considerations
 might rule these out (no conventional stars?
 no inflation?  no structure?)  I don't think that most of us
 would find
this Minimalist Anthropic Principle particularly disturbing.
It would not render the theory unpredictive or unfalsibiable.

What about theories with less supersymmetry?  In some sense, 
theories with $N=1$ supersymmetry are not, at first sight, 
obviously different than those with $N=0$ supersymmetry.  After 
all, generically one expects potentials on the moduli spaces of 
both albeit perturbatively in one case and
non-perturbatively in the other.
In my view, there is some tentative
evidence that there is a real distinction. First, 
however, since we are speaking, at best, of approximate 
supersymmetry, we need some definitions.  I will refer to $N=0$ 
theories as theories for which there is no limit of the moduli 
space with only a finite number of supersymmetries. Examples 
include (compactifications of) the $O(16)\times O(16)$ heterotic 
theory, the non-supersymmetric heterotic M-theories studied by 
Fabinger and Horava, and others. $N=1$ theories are those with 
only four supersymmetries in {\it some} limit of the moduli 
space.  Familiar example include the heterotic and Type I 
theories on Calabi-Yau spaces, orientifolds of Type II theories on 
Calabi-Yau, and many more.

The $N=0$ theories have a number of properties, which suggest 
that they might be qualitatively different from the supersymmetric 
ones.  Conceivably many (most?) of them suffer from various 
diseases. One potential problem is the appearance of tachyons 
appear in the moduli space, but it is not clear that this signals 
any kind of consistency.  Fabinger and Horava\cite{fabingerhorava} 
have argued that $N=0$ theories may suffer, generically, from other 
instabilities, particularly Witten's ``decay to 
nothing".\cite{bubbleofnothing}. With Fox and 
Gorbatov,\cite{dinefoxgorbatov} I have been following up this 
latter issue.  We have sharpened, but nor resolved, the question 
of the extent to which bubbles of nothing are generic to these 
spaces.  There are, we believe, good reasons to think these 
instabilities {\it are} generic.  We have also understood how these 
instabilities arise even in the presence of a potential on the 
moduli space, in $N=0$ theories (but not in $N=1$ theories).
But while the prospect of decay to nothing is 
frightening, it is not clear whether this means that these states 
don't make sense as quantum theories. Still, it seems possible 
that with some work we can show that large classes of $N=0$ 
theories don't make sense.

The discretuum suggests another approach to this question. As we 
will see, it is plausible ({\it but certainly not established}) 
that vacua with low energy supersymmetry, consistent with modest 
anthropic constraints, are by far the most numerous.

Before exploring anthropic arguments, it is worth noting that in 
controlled approximations, we don't presently know how to make 
sense of most (any?) string solutions with four or less 
supersymmetries.   Generally, because there are potentials
on on the moduli space, they are cosmological with 
singularities in their past or future, which we don't know how to 
treat.\cite{bdcosmo} Perhaps this is cause for optimism.  After 
all, for space-like singularities, we know that string theory 
sometimes provides a resolution, and we know (in weakly coupled 
strings and sometimes elsewhere) the criterion (modular 
invariance). Perhaps there is an analogous condition for 
time-like singularities. If we were lucky, perhaps this condition 
would yield only a small set of consistent theories.

\section{The Dreaded Anthropic Principle}

Linde was probably the first to realize that inflation leads to a 
framework in which one might sensibly implement the anthropic 
principle.\cite{lindeanthropic} Perhaps in a very vast universe, 
the fundamental parameters take different values in different 
regions.  The most promising application of this idea has long 
been to the problem of the cosmological constant, for which it is 
fair to say that we currently have no equally successful candidate 
solution.\cite{lambdaanthropic}

The response to this suggestion has been enthusiasm in some 
quarters and revulsion in others. I have been more agnostic, and 
I strongly believe that the question itself is a scientific one.  
If string theory provides a means to implement the anthropic 
principle, then we {\it have no choice but to consider it 
seriously.}  Rather than wasting time arguing about 
philosophical issues, I think the smart money will be on trying 
to figure out whether one can make predictions or falsify the 
theory. Some ideas for how predictions might emerge will be 
discussed below. If string theory does not provide a setting for 
the anthropic principle, while we can't, perhaps, ``disprove" it, 
I think we can justifiably, {\it and smugly}, ignore it.

There are, two my knowledge, two principle scenarios for 
implementing the anthropic principle:
\begin{enumerate}
\item Extremely light 
scalars:\cite{bankslight,lindelight,vilenkinlight}  Here, if
one is to solve the cosmological constant problem, one 
requires a scalar with a mass smaller than the present Hubble 
constant, but which varies over such a large region of field 
space that it can cancel an energy density at least of order 
TeV$^4$. \item A discretuum or landscape
\cite{banksdineseiberg,bp,sethietal,kklt} (as for the references, the first provided a 
simple model based on an ``irrational axion", but no example of this
type has been exhibited in string theory; the second and third used fluxes in a manner 
not too different than that currently of interest; the last is 
the most fully developed of these ideas).
\end{enumerate}

Each of these scenarios is quite radical, and their plausibility 
cannot be assessed without a theory which is fundamental (in the 
sense of including gravity and other interactions, and finite) 
like string theory.

\subsection{Extremely Light Scalars}

In string theory, it appears unlikely that there are scalars 
with the requisite properties to solve the
cosmological constant problem.\footnote{Dimopoulos and 
Thomas\cite{dt} have recently suggested a possibility which could 
conceivably be realized in string theory: they have argued that a 
conformal field theory might yield a huge $Z$ factor for some 
scalar field.  This possibility is worthy of further 
exploration.  This discussion itself is an example of how the 
anthropic principle is an issue whose validity and relevance can 
be decided by scientific means.} Ordinary scalars in string 
theory have masses generally have masses consistent with
 dimensional analysis, e.g. set by the scale of supersymmetry
 breaking.  String theory also has periodic scalars,
 for which shift symmetries can suppress masses.  So these might
 be candidates, but one needs decay constants far greater (exponentially
 greater) than the Planck mass.  Searches in string/M theory have not
 yielded any candidates.\cite{bdfg}  (Such fields would
 also be of interest as candidate inflatons.\cite{lisanima})

While this cannot yet be considered a theorem, it seems unlikely 
that this sort of
  implementation of the anthropic principle is realized
 in string/M theory.   What is perhaps most interesting is that it
 might be possible, with a finite amount of effort, to rule out this implementation
 completely.

 \subsection{Discretuum or Landscape}

There have been various proposals for a discretuum.  That of 
\cite{banksdineseiberg} does not seem to be realized in string 
theory.  That of \cite{bp}, while interesting, made assumptions 
about the stabilization of moduli for which there was no 
support.  The proposal of \cite{kklt} is the most promising to 
date, including, as it does, a detailed picture of the 
stabilization of moduli.
This proposal is the subject of the next section.

\section{The Proposal of KKLT}

KKLT have put forth a quite detailed proposal for how a landscape 
might emerge in string theory. There is not space here to fully 
review this set of ideas, but we should note, first, that they 
are based on effective field theory.  Banks has argued 
cogently that one cannot use effective field theory to study 
multiple vacua in theories of gravity.\cite{bankseffectiveaction}
 For example, in many circumstances there
are no transitions between the different states, and an observer 
in one can not do experiments which will indicate the existence 
of others. So it is not clear
that the multiplicity of states has any meaning.
These arguments have been reviewed and elaborated in \cite{bdg}.

Setting aside these larger questions of principle, I want to 
focus here on questions of self-consistency with
the effective action analysis. I will not be able 
to fully answer whether KKLT have actually demonstrated the 
existence of a discretuum.  My tentative conclusion will be that 
there is a discretuum of states with $N=1$ supersymmetry, but 
that only a limited set of quantities can actually be computed in 
these states.  Because of the latter problem, studies of 
supersymmetry breaking may be more problematic, but the existence 
of a supersymmetric discretuum is at least plausible.  Some 
of these questions can probably be answered through further 
investigation.

Compactifications of string theory (IIB on a CY, X, or $F$-theory 
on a Calabi-Yau four-fold, for
 definiteness) permit many possible quantized fluxes,
 $F_{IJK}, H_{IJK}$.
The number of possible three cycles ($b_3$) can be of order 
$100$'s.  These fluxes are
 not highly constrained.  Tadpole cancellation gives
 one condition on many fluxes.  Because there are
 so many possible choices of fluxes, there are potentially
 an exponentially large number of states.

KKLT, following earlier work\cite{gkp,becker,acharya} 
noted that the presence of flux tends to stabilize moduli. For 
example, \cite{gkp} considered orientifolds of Type II theory on 
a Calabi-Yau space near a conifold point.  If $z$ is the modulus 
describing the distance from the point, then in the presence of 
flux on the collapsing three cycles one finds both stabilization 
and warping.  There is a superpotential for $z$ and the dilaton, 
$\tau={i \over g} + a$, \beq W= (2 \pi)^3 \alpha^\prime (M {\mathcal 
G}(z) -K \tau z) \eeq where $M$ and $K$ denote the flux quanta, 
and \beq {\mathcal G}(z)= {z \over 2 \pi i}\ln(z) + 
{\rm~holomorphic}. \eeq

This effective action has a supersymmetric minimum where \beq D_z 
W = {\partial W \over \partial z}+{\partial K \over \partial z}W 
=0 \eeq This is solved by: \beq z \approx {\rm exp}(-{2\pi K \over M 
g_s}) \eeq If the ratio $N/M$ is large, then $z$ is very small.  
The corresponding space can be shown to be highly warped.
In addidtion, 
 \beq 
W_o = <W> \eeq is exponentially small.

Including additional fluxes, it is possible to fix other complex 
structure moduli and also the dilaton, $\tau$. \beq W= (2 \pi)^3 
\alpha^\prime[M {\mathcal G}(z)-\tau (Kz + K^\prime f(z))] \eeq
$$D_\tau W = {\partial W
\over \partial \tau} + {\partial K \over \partial \tau} W =0$$ for
$$\bar \tau = {M {\mathcal G}(0) \over K^\prime f(0)}~~~~~W= 2(2 \pi)^3 \alpha^\prime
M {\mathcal G}(0) $$ $z$ is still exponentially small, and the space 
is highly warped, but {\it now $W_o$ is no longer exponentially 
small}.  This is crucial to the KKLT picture.
In the limit that $R$ is very large, there is no 
potential for $R$; the compactification radius is not fixed.

\subsection{Fixing the Remaining Moduli?}

KKLT noted that in flux vacua, $W_o$ is generically large (of 
order
 some typical flux integer), but argued that among the
 vast number of possible fluxes, $W_o$ will sometimes be small,
 simply by chance.
Other effects (e.g. gluino condensation) will generate a 
superpotential for $\rho = R^4 + i b$: \beq
                 W= W_o + e^{-\rho/c}
                 \eeq
This has a supersymmetric minimum, with \beq
                \rho \sim  -c \ln(W_o)
                \eeq
Consistency of the analysis requires that $\rho$ be large, but 
this will be the case only in an exponentially small fraction of 
states.

If there is a systematic approximation, it consists of 
integrating out
 the string modes followed by the KK modes, the complex structure and dilaton,
 and finally the
 radial mode.  Consistency requires a hierarchy of masses:
\beq M_{string}^2 >> M_{kk}^2 = {1/R^2} >> M_{\tau^2,z^2} >> 
m_\rho^2. 
\label{hierarchy}
\eeq 
Before making detailed calculations, these conditions seem to hold
at large radius.  The masses of the $\tau$ and $z$ fields are suppressed
by $\rho$ relative to the Kaluza-Klein modes, and by $\rho^{3/2}$ relative
to the string scale.  But in the presence of large fluxes or large $b_3$, there
are a number of sources of enhancement, which
can be inferred from the expression for the superpotential:
\begin{enumerate}
\item  Factors of fluxes,  $(N,M)^2$ \item  Factors of $\tau$, 
$\tau^2$ \item  Factors due to the large size of the mass matrix, 
$b_3$.  In the absence of any detailed understanding of the mass
matrix, we might worry that a large, random matrix has eigenvalues which
grow with the size of the matrix.
\end{enumerate}
This list suggest that one might expect that \beq {m_{z,\tau}^2 \over 
M_{string}^2} = {b_3 \tau^2 N^2 \over \rho^{3/2}}. \eeq

If we require, say $\tau =5$, and suppose that a typical $N$ is 
of order $3$, this requires that $\rho$, be quite large, and 
therefore that $W_o$ is {\it extremely} small.  Indeed, if we 
take for the number of states that suggested by KKLT and by 
Douglas, $e^{b_3 \ln({\chi \over 24}/2)}$, there might well not 
be enough states that one could imagine doing a self-consistent 
computation in {\it any} of them.

But this may not call the KKLT discretuum into 
question.\footnote{I thank the participants at the KITP workshop 
on string cosmology for discussions of these issues, and 
especially Joe Polchinski.}  Imagine studying the theory at 
extremely large $\rho$.  Here, one can compute $W_o$ reliably.  
One can also compute the leading $\rho$-dependent terms in the 
superpotential. Because of non-renormalization theorems, this 
form of the superpotential will remain valid until the point 
where higher order exponentials become important. This
requires only that $e^{\rho}$ be small, which would seem to be a
much weaker requirement than equation \cite{hierarchy}.

Supersymmetry breaking, from this point of view, could be more 
problematic. One will not have the same level of control of the 
Kahler potential, for example. 

It is worth stressing what serves as the small parameter which 
justifies analysis in the landscape.  It is not large flux 
numbers or large $b_3$, per se, but the fact that in the vast 
discretuum of states, there is a small fraction -- but large 
number -- of states with small $W_o$.

\section{Supersymmetry and Supersymmetry Breaking}

For now, let's accept that a discretuum exists, and that the 
universe samples all of these
   states in its cosmic history.
   KKLT argued that  anti D3-branes, located well down the throat can give exponentially
       small effects in a warped geometry.
An alternative possibility, which is perhaps somewhat easier to 
think about, is to note that in some fraction of this vast array 
of states, the low energy dynamics presumably breaks supersymmetry.
(By assumption, the discretuum contains states with complicated
gauge groups and chiral fermions.)  Then
   \beq
         V = {\rm exp}(-8\pi^2/b_o g^2)
         \eeq
       If $g^{-2}$ is distributed more or less uniformly, V will be distributed roughly
         uniformly on a log scale.  Correspondingly, there will be a substantial number
         of states where the cosmological constant,
     \beq
     V = [{\rm exp}(-8 \pi^2/b_o g^2)-3\vert W_o\vert^2]
\eeq is small compared to $\vert W_o\vert^2$.

\section{Anthropics}

So we have a picture in which there are many, many states.  Among 
these states, quantities relevant to low energy physics vary:
\begin{itemize}
\item Low energy gauge groups \item The matter content \item The 
values of the parameters of the low energy lagrangian.
\end{itemize}

If the universe samples all of these states, there will only be 
observers in a subset with suitable properties.

\subsection{The Cosmological Constant}

The cosmological constant provides the most compelling 
application of the anthropic principle.\cite{lambdaanthropic}  
Holding all other fundamental and cosmological parameters fixed, one 
finds that suitable structure forms only if $\Lambda$ is less then 
or of order $10$ times its observed value\footnote{Allowing other 
quantities to vary permits a far broader range of $\Lambda$.  
See, e.g., \cite{aguirre}}.

\subsection{Does the Anthropic Landscape Predict Low Energy Supersymmetry}

We cannot prove at present that the landscape predicts low energy 
supersymmetry, but it seems likely that it would.  As Douglas has 
explained (see also \cite{bdg}), for small $W_o$, the distribution 
of $W_o$ is likely to be uniform. Suppose, also, that the origin 
of supersymmetry breaking is dynamical, as described above, and 
that $8 \pi^2 \over g^2$ roughly uniformly distributed.  Then, 
e.g., if $W_o = 10^{-28}$ (susy breaking $\sim 10^4$ GeV), in 
about $10^{-3}$ of states,
$$V_o = e^{-{8\pi^2 \over b g^2}} - 3 \vert W_o\vert^2 < \vert W_o \vert^2.$$
In this subset of states one has the possibility of obtaining a 
small cosmological constant.

Suppose that the anthropic argument for $\Lambda$ is correct.  We 
can compare the number of states with sufficiently small $\Lambda$ with 
and without supersymmetry. Of course, since we have not done a 
reliable counting, we are describing here only a program which 
might lead to such a prediction, and its hypothetical results.

Without supersymmetry, we might expect simply:
$$P(\Lambda) \approx {\Lambda \over M_p^4}.$$
(here $P$ is the probability of cosmological constant less than 
$\Lambda$).

With supersymmetry, we start with the probability of small $W_o$:
$$P(W_o) \approx {W_o \over M_p^3}.$$
Now small $\Lambda$ will favor small supersymmetry breaking.  This is
the usual argument about the connection of the cosmological constant
and the scale of supersymmetry breaking, and it remains valid in
this framework. This means that 
we will require more anthropic input. In particular, we 
might imagine that the ratio: $M_w/M_p$  is fixed (note that we 
would also require something like this for non-supersymmetric 
theories, presumably, to account for the hierarchy).  So we might 
estimate the fraction of suitable states along these lines:
\begin{enumerate}
\item $10^{-10}$  have suitable susy breaking (i.e. low energy 
gauge groups with suitable properties) \item $10^{-2}$ have susy 
breaking comparable to $W_o$ \item  $10^{-13}$ have suitable 
$W_o$. \item $10^{-60}$ of these have small $\Lambda$.
\end{enumerate}

So we guess that for supersymmetric states, we pay a factor of 
$10^{-85}$ to realize this set of anthropic constraints,  vs. 
$10^{-120} \times 10^{-32}$ for non-susy states. So SUSY wins 
unless there are an overwhelmingly large number
 of non-susy states relative to supersymmetric ones.
Note that this picture favors susy breaking at the lowest
  possible scale (gauge mediation?)

  So one sees here how the anthropic principle, coupled
  with knowledge about the distribution of states,
  might lead to a real prediction.   But
   before getting too excited, there are other issues to face
   in the flux discretuum.

\subsection{Anthropic Pitfalls}

A program of implementing the anthropic principle within the 
landscape faces several hurdles, and at least at first sight, 
seems likely to fail.  Within the landscape, as we currently 
imagine it, there are numerous states with different gauge 
groups, particle content, and couplings.  These features must 
either be fixed anthropically, or are otherwise random.  But 
there are many features of the Standard Model which seem neither 
anthropically constrained {\it nor random}.\cite{donoghue}

In thinking about this sort of anthropic selection, it is useful 
to organize the problem by considering physics first at very 
large distance scales, and then at progressively shorter scales, 
using the language
 of effective actions and the renormalization group.

At the very largest distances, we face the problem of the 
cosmological constant, which we have already discussed.  At 
shorter distances, we face the question of the existence of an 
unbroken $U(1)$ symmetry. This is plausibly anthropic.  The 
question of whether the strong group is $SU(3)$ or something else 
might be anthropic, but this is more difficult to decide.  E.g. for groups 
other than $SU(3)$, by varying $m_u$ and $m_d$ we can probably 
reproduce many features of nuclear physics.  Conceivably 
deuterium is essential for stellar nucleosynthesis, and this might 
single out $SU(3)$.  The ratio $m_e/\Lambda_{qcd}$, another 
important parameter of the low energy lagrangian, might be fixed, 
for example, by molecular physics and/or by astrophysics.  The 
relative size of $m_u$ and $m_d$ might also be determined by the 
details of nuclear physics; at the grossest level, the fact that 
$m_d>m_u$ is necessary for proton stability (as opposed to 
neutron stability).

While plausibly constrained by anthropic arguments, making 
persuasive arguments for these quantities will be challenging, 
at the very least.  But at higher energies, we encounter 
couplings whose prediction is more problematic:  $m_s, m_c, m_b, 
V_{km}$.  It is hard to see how these could be anthropic (if 
there were obvious anthropic arguments, one might have predicted 
the values of these quantities prior to their discovery). One of 
the most puzzling is $\theta_{qcd}$.  It is not at all clear what 
sort of anthropic argument might require $\theta_{qcd} << 1$.

In the landscape, it would appear that all of these quantities 
are simply random numbers.  $\theta_{qcd}$, for example, is 
presumably not small in a typical vacuum.  The fluxes, 
generically, break CP and contribute to $\theta_{qcd}$.
Moreover, the assumption is that all of 
the moduli are fixed would seem to preclude axions.  E.g. cross 
terms in the potential for $\rho$ involving $W_o$ will give a 
large potential to the pseudoscalar component.

At still high energies, one encounters further questions.  The 
weak gauge group may be hard to understand anthropically.  Dark matter,
cosmological parameters (the size of 
inflationary fluctuations, the number of $e$-foldings, and the 
like) will pose deeper challenges.

In the flux discretuum, the parameters of low energy physics seem to be random
numbers.
If this is really true,
the landscape is not a correct description of physics. 
Alternatively, there are some set of principles in the
landscape which explain those laws of nature
which do not seem to be anthropically constrained.
 Within the flux discretuum, it is not obvious what these might be.
  As another example, anthropically, the
  proton lifetime is probably not required to be much
  larger than $10^{16}$ years.  So one might to hope to understand
  the length of the proton lifetime from symmetries.
  But  most states of the flux discretuum
   don't have symmetries.\footnote{For example,
   in the case of the $T_6/Z_2$ orbifold,
   one has a $Z_2^5 \times S_6$ symmetry
   at some points in the moduli space.  But half of the fluxes
   must vanish to preserve even one $Z_2$.  In more realistic
   models, this is likely to correspond to a drastic
   reduction of the number of states in the discretuum.}
     For $\theta_{qcd}$, we might want to find a
   reason why some modulus is not fixed at high energies.
One could imagine that this is cosmological.

\section{Conclusions}

Whether one likes it or not, it is quite possible that quantum 
theories of gravity predict a landscape. 
It is plausible that the flux vacua explored by KKLT exist, but 
it is by no means certain.  Perhaps the strongest reason for 
doubt is Banks' general critique of effective action methods in 
gravity theories, but I have mentioned
some concerns within the conventional
framework of effective actions.  I have also mentioned some curious issues of 
scales, within the framework of effective field theories.  One 
interesting feature of the proposed discretuum which we have 
noted is that the small parameter is essentially a random 
variable.

If there is a discretuum, it seems quite likely that we will be able to 
falsify this whole picture. Typical states, even subject to 
anthropic constraints, will disagree violently with observation.  
To avoid this conclusions, we would need to show
that in 
some subset of states -- which are picked out by other 
considerations -- there is some rational explanation of the many 
features of the Standard Model which don't seem susceptible to 
anthropic explanation.  We have seen, for example, how low energy 
supersymmetry could conceivably emerge from the requirement of 
small cosmological constant and suitable electroweak symmetry 
breaking. Perhaps some considerations might lead to approximate 
flavor symmetries, small $\theta$, and the like.  On a pessimistic
note, we have argued that symmetries are unlikely to play this role.

Mention of the anthropic principle brings out strong reactions 
from most physicists, who ask what can be the role
of science in such a situation.  But the lesson of the KKLT 
proposal is not so pessimistic.  First, the existence of a 
landscape within string theory is a question we should be able to 
decide.  If we decide that there is such a discretuum, we will 
probably be forced to contemplate the anthropic principle; if not, we
can dismiss it.  But even if we do adopt the anthropic principle,
it will at best explain only a few quantities: 
either we will falsify string theory, or we will uncover 
principles which explain most of the features of the Standard 
Model.  We will likely make additional predictions for 
accelerators and cosmology as well. So surrender to the anthropic 
principle will not be necessary or possible; we won't have to 
give up. 



\section*{Acknowledgments}

I wish to thank my collaborators on this work, Tom Banks and Elie 
Gorbatov.  I also wish to thank Michael Douglas, Patrick
Fox,  Renata Kallosh,
Shamit Kachru, Andrei Linde, Joe Polchinski, Eva Silverstein, and Sandip Trivedi
for conversations which have shaped my thinking on these matters.
I particularly want to thank the hospitality of the KITP and
the participants in the string cosmology workshop for helpful
comments and criticisms.



\end{document}